\documentclass{emulateapj}
\usepackage{apjfonts}
\usepackage{epsfig}
\usepackage{natbib}

\newlength{\colwidth}

\newcommand{\be}{\begin{equation}}
\newcommand{\ee}{\end{equation}}
\newcommand{\bee}{\begin{eqnarray}}
\newcommand{\eee}{\end{eqnarray}}

\newcommand{\msol}{\hbox{${\rm M}_\odot$}}

\newcommand{\tvir}{T_{\rm vir}}
\newcommand{\mthresh}{M_{\rm t}}
\newcommand{\nsats}{n_{\rm sats}}
\newcommand{\vmax}{v_{\rm max}}
\newcommand{\zreion}{z_{\rm reion}}
\newcommand{\LCDM}{$\Lambda$CDM }

\newcommand{\K}{{\rm K}}
\newcommand{\VL}{{\em via Lactea\ II}}

\newcommand{\hinv}{h^{-1}}

\newcommand{\lsim}{\lower.5ex\hbox{\ltsima}}
\newcommand{\gsim}{\lower.5ex\hbox{\gtsima}}
\newcommand{\ltsima}{$\; \buildrel < \over \sim \;$}
\newcommand{\gtsima}{$\; \buildrel > \over \sim \;$}

\usepackage{color}

\newcommand{\addition}[1]{\textcolor{green}}

\shortauthors{BUSHA ET AL.}
\shorttitle{Inhomogeneous Reionization and Satellites Galaxies}
\begin{document}

\title{The Impact of Inhomogeneous Reionization on the \\
Satellite Galaxy Population of the Milky Way}

\author{Michael T. Busha, Marcelo A. Alvarez, Risa H. Wechsler, Tom
Abel, Louis E. Strigari}

\affil{Kavli Institute for Particle Astrophysics and Cosmology \\
Stanford University, Stanford, CA, 94305\\
SLAC National Accelerator Laboratory, Menlo Park, CA, 94025\\
mbusha, malvarez, rwechsler, tabel, strigari@stanford.edu}

\begin{abstract}

  We use the publicly available subhalo catalogs from the \VL
  ~simulation along with a Gpc-scale N-body simulation to understand
  the impact of inhomogeneous reionization on the satellite galaxy
  population of the Milky Way.  The large-volume simulation is
  combined with a model for reionization that allows us to predict the
  distribution of reionization times for Milky Way mass halos.
  Motivated by this distribution, we identify candidate satellite
  galaxies in the simulation by requiring that any subhalo must grow
  above a specified mass threshold before it is reionized; after this
  time the photoionizing background will suppress both the formation
  of stars and the accretion of gas.  We show that varying the
  reionization time over the range expected for Milky Way mass halos
  can change the number of satellite galaxies by roughly two orders of
  magnitude.  This conclusion is in contradiction with a number of
  studies in the literature, and we conclude that this is a result of
  inconsistent application of the results of \cite{Gnedin00}; subtle
  changes in the assumptions about how reionization affects star
  formation in small galaxies can lead to large changes in the effect of
  changing the reionization time on the number of satellites.  We
  compare our satellite galaxies to observations using both abundance
  matching and stellar population synthesis methods to assign
  luminosities to our subhalos and account for observational
  completeness effects.  Additionally, if we assume that the mass
  threshold is set by the virial temperature $\tvir = 8 \times 10^3\K$
  we find that our model accurately matches the $\vmax$ distribution,
  radial distribution, and luminosity function of observed Milky Way
  satellites for a reionization time $\zreion = 8^{+3}_{-2}$,
  assuming that the \VL ~subhalo distribution is representative of the
  Milky Way.  This results in the presence of $540^{+100}_{-340}$
  satellite galaxies.

\end{abstract}


\section{INTRODUCTION}

While the cold dark matter (CDM) paradigm has been very successful in
explaining the large scale distribution of matter in the universe, one
final test lies in its ability to predict the distribution of matter
on small scales, including the distribution of satellite galaxies
around the Milky Way.  In the hierarchical model of structure
formation, massive objects such as the Milky Way halo are built up
through a series of mergers where small, dense objects collapse early
and merge to form larger objects.  High-resolution numerical simulations have shown 
that the
dense cores from a significant number of these small building blocks
should survive today as gravitationally bound subhalos
~\citep{Klypin99,Moore99b,Diemand08,Springel08}.   From these 
initial results~\citep{Klypin99,Moore99b} and from semi-analytic
modeling~\citep{Kauffmann93} it
was also clear that there are far fewer luminous dwarf satellites
around the Milky Way than bound dark matter subhalos in the 
simulations. 

There has been no shortage of solutions proposed to explain this discrepancy,
including those which modify the dark matter power spectrum to reduce the amount of
small scale power through warm dark matter-like models
\citep[e.g.,][]{Colin00,Bode01,Zentner03,Busha07}.  Less exotic models, however,
such as the presence of a number of ``dark-dark halos,'' dark matter
subhalos that do not host galaxies, provide a well-motivated resolution to this
issue within the CDM paradigm.  In particular, the presence of
photoionizing radiation is expected to have a significant effect on
the ability of a subhalo to host a luminous galaxy~\citep{Bullock00, Gnedin00,
  Benson02,Somerville02, Dekel03a,Shaviv03,Li08,Madau08b, Koposov09}.
Star formation in dark matter halos in the early universe is only
expected to be able to begin once the halo has grown massive enough to
cool efficiently by atomic cooling, typically around $\tvir \sim
10^4~K$.  However, before many halos are able to reach this mass, the
universe enters the phase of reionization, in which photoionizing UV
radiation is released by the early generations of stars and quasars.
This ionizing radiation heats the halo gas to a temperature of a few
times $ 10^4\K$, preventing it from being pulled into the shallow
potential wells of halos with virial temperatures lower than $\sim
10^5~K$ \citep{Thoul96,Dijkstra04}, effectively
suppressing further star formation.

Recent analysis of the Sloan Digital Sky Survey (SDSS) has resulted in
the discovery of a large number of low-surface brightness dwarf
galaxies~\citep{Willman05a, Willman05b, Belokurov06,   Zucker06b,
Zucker06a, Belokurov07, Irwin07, Walsh07, Belokurov08}, which is now
beginning to shed a new light on the mapping of galaxies onto dark
matter halos at the low-luminosity end. Several of these
newly-discovered satellites  have luminosities similar to those of the
least luminous globular clusters, and a dynamical analysis indicates
that they have the largest mass-to-light ratio of any known
galaxies~\citep{Martin07,Simon07,Strigari08,Geha08}.  Given both the
magnitude limit and sky-coverage fraction of the SDSS survey, it is
certainly reasonable to assume that  we have only detected a fraction
of the Milky Way satellites~\citep{Koposov08,Tollerud08,Walsh09}, and there are
exiting prospects for discovery of more satellites in future deep and
wide surveys ~\citep{Abbott05,Keller07,Ivezic08}. A full understanding
of the mapping between luminous satellite galaxies and dark matter
subhalos will require a measurement of the luminosity distribution,
radial distribution~\citep{Kravtsov04b,Willman04}, and the kinematic
properties of the satellites~\citep{Strigari07c}.  

Recently, it has become understood that the universe actually
reionizes quite inhomogeneously, even on very large scales \citep[100
Mpc$-$1 Gpc; e.g.,][]{Sokasian03,Barkana04,Iliev06,Lidz07, Alvarez08}.  Given
constraints on the global reionization history, e.g.~from the cosmic
microwave background polarization \citep[e.g.,][]{WMAP3}, there is
still a significant uncertainty in the precise reionization epoch of
the matter in the Milky Way.  In particular, the calculations of
\cite{Alvarez08} indicate that there is substantial scatter in the
reionization histories of halos of a given mass, and that on
average Milky Way mass halos reionization redshifts have approximately
the same distribution as that of the universe as a whole.  
A change in the reionization time of the Milky Way may have a dramatic
impact on its satellite population because the reionization history
may directly affect the ability of a subhalo to reach the size where gas
is able to cool and begin the star formation process.  The primary aim
of this paper is to understand what, if any, effect the reionization
epoch of a given Milky Way halo has on its satellite population.

The effect of the reionization epoch on the satellite population has been
previously addressed in the literature, with most studies finding
little effect \citep[e.g.,][]{Somerville02,Kravtsov04b}.  These
studies primarily have addressed the luminosity range of the
``classical dwarfs'' in the Milky Way.  Given the dramatically
different observational picture that has emerged with new observations
from SDSS, combined with the possible importance of a spread in
reionization epochs expected from inhomogeneous reionization, we
re-investigate this question here.  We combine a high-resolution dark
matter simulation \citep[the \VL ~simulation of][]{Diemand07b, Diemand08}
with various assumptions about star formation in small halos, and
compare to up-to-date constraints from the full observed satellite
galaxy population.  In this work, we critically examine the
assumptions about the rate at which photoionizing UV radiation is able
to heat halo gas.  We find that the exact time a halo reionizes can
have a significant impact on the satellite population, and use
comparisons with the Milky Way's satellite distribution to constrain
the reionization time of our own halo.


In section \S2 we discuss our simulation and models, including methods
for identifying subhalos that host satellite galaxies, and determining
magnitudes of these galaxies.  In \S3 we discuss the observational
sample of satellite galaxies that we compare to our model, including
measurements and corrections for the luminosity, $\vmax$ and radial
distribution functions.  In \S4 we directly compare our modeled and
observed samples in a manner that fairly accounts for the
incompleteness of the observations.  In \S5 we discuss how our results
compare with previous work in the literature, in particular addressing
the differences between our results and those that have found little
change in the satellite population for varying reionization times
\citep{Somerville02, Kravtsov04b}.  Finally, we present our
conclusions in \S6.  

\section{SIMULATIONS AND MODELING}

\subsection{Modeling the Large-scale Structure of Reionization}

In order to understand the distribution of reionization times of Milky
Way sized galaxies, including any effect this may have on the
satellite galaxy population, it is necessary to understand the
distribution of dark matter on the largest scales.  We use the recent
results of \citet{Alvarez08}, in which the reionization process was
modeled using an N-body simulation of a 1 Gpc~$h^{-1}$ box combined with an
analytic prescription for predicting the reionization time for all
points in the box, as described below. For more details on the
reionization simulation and the halo correlation, see
\citet{Alvarez08}.

The N-body simulation used the code Gadget2 \citep{Gadget2} to evolve
$1120^3$ dark matter 
particles in a cosmology with $\Omega_m = 0.25$, $\Omega_{\Lambda} =
0.75$, and $\sigma_8 = 0.8$, with particle mass 
resolution $M_p = 4.94\times 10^{10} \hinv\msol$.  The initial conditions
were generated using the 2nd order Lagrangian perturbation code 2LPT
\citep{Crocce06}.  A Friends-of-Friends group finder based on the
Ntropy framework \citep{Gardner07} was run on the $z=0$ output and
identified all halos with $M_{FoF} \ge 1.58\times 10^{12}\hinv\msol
= 32$ particles.  This simulation was run in conjunction with the
LasDamas
collaboration\footnote{http://lss.phy.vanderbilt.edu/lasdamas/}.
The reionization history for this simulation was 
then calculated using the density field of the initial conditions.
Working at a single 
point in space, the dark matter density field is smoothed over a
series increasing radii and use the EPS formalism to ask, for a given
redshift, what is the smallest (if any) radius at which the fraction
of collapsed mass in halos greater than some threshold is greater than
a specified fraction, $1/\zeta_c$. The reionization time, $\zreion$,
for that point is the earliest redshift at which this criteria is
first met for any radius. For this simulation we use $\zeta_c = 10$
and a threshold mass $10^8 \msol$.  Combined with the group catalog,
this results in a reionization time for each halo in the simulation.

\subsection{Modeling Galaxy Formation \label{sec:galform}}

While the above simulations are sufficient for measuring the
reionization history of Milky Way mass halos, significantly higher
mass resolution is needed to understand the subhalo distribution.  For
this, we use the publicly available mass accretion
histories\footnote{http://www.ucolick.org/$\sim$diemand/vl/data.html} from
the high-resolution \VL ~simulation \citep{Diemand07a, Diemand07b, Diemand08}.
This simulation models a single dark matter halo with virial 
mass $M_{halo} = 1.9 \times 10^{12} \hinv\msol$ that is able to
resolve subhalos down to mass limit $M_{sub} > 10^6 \hinv \msol$.
The publicly available data includes the distribution, tidal mass, and
$\vmax$ histories for the most massive progenitors of all $z = 0$
subhalos back to redshift $z = 28$.  

In order to connect the \VL ~dark matter subhalo population to a
satellite galaxy population, we assume that stars begin forming once
atomic cooling becomes effective, when a halo shock heats to virial
temperature $\tvir \approx 8\times 10^3K$, but that reionization heats
the gas in the subhalos to the point where this becomes ineffective
\citep{Thoul96, Kepner99, Wise08}.  We treat this heating as an
instantaneous process (see section 5 for a discussion), causing
reionization to end star formation for the vast
majority of the $\sim 2500$ potential satellite galaxies; for these
low-mass halos, all star formation must happen before $\zreion$.  With
this in mind, we can define a subhalo as being a satellite galaxy
using a two parameter model: A subhalo must grow to a threshold mass,
$\mthresh$, above which HI cooling will allow star formation, before
the host halo reionizes at $\zreion$ in order to host a satellite.

While we demonstrate the effects of varying both parameters in the
next section, the work of \cite{Abel02} uses high resolution AMR
simulations to model the formation of the first stars and indicates
that we anticipate $\mthresh \approx 10^6 - 10^7 \hinv\msol$. It is
important to note that this process of hydrogen cooling simply defines
a minimum mass of the population of the dark matter subhalos that
could host satellite galaxies. However, this work predicts the stars
forming in these halos to be very massive and short--lived. As such
these very first star forming halos cannot be the direct progenitors
of Milky Way satellites, which are observed to be metal-enriched
objects with stars presumably of masses less than a solar mass. More
relevant here are the calculations of \cite{Wise08}, who followed the
build up of halos up to the masses when they start cooling via
Lyman-alpha from neutral hydrogen. They included the radiative as well
as the supernova feedback from the first generation of massive
stars. The short-lived sources keep ionizing the baryonic material in
the halos they form in, as well as their surroundings. However, as
they turn off, material can cool again and repopulate the dark matter
halos. So while the baryon fraction \citep[Fig. 4 in][]{Wise08}
fluctuates and decreases at times to as little as 10\%, star formation
can continue as long as no sustained external UV flux sterilizes the
halo. The latter case severely limits star formation and has been
discussed many time in the literature
\citep[e.g.,][]{Babul92,Thoul96,Kepner99, Dijkstra04}. It seems clear
then from the limited guidance we have from numerical simulations that
most Milky Way satellite halo progenitors experienced most
of their star formation before they are permanently ionized.

\begin{figure}
\plotone{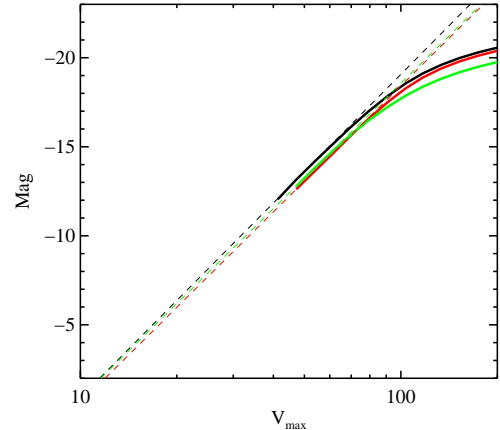}
\figcaption{
The relationship between magnitude and $\vmax$ for the $r-$, $g-$, and
$V-$ bands using abundance matching (solid red, green and black
lines).  The dashed lines show power law fits to the low-luminosity
end.  
\label{fig:abundance}
}
\end{figure}

Once we have identified satellite galaxies in the simulation, we must
assign magnitudes to them in order to make direct comparisons with
observations and to account for observational completeness effects.
This is done using two methods.  First, we use a halo abundance
matching method \citep{Kravtsov04a, Blanton08}.  Here, luminosities
are assigned to halos by assuming a one-to-one correspondence between
$n(<M_V)$, the observed number density of galaxies brighter than
$M_v$, with $n(>\vmax)$, the number density of simulated halos with
maximum circular velocities larger than $\vmax$.  For the distribution
of magnitudes, we use the double-Schechter fit of \cite{Blanton05} for
low luminosity SDSS galaxies in the $g-$ and $r-$bands down to $M_r =
-12.375$.  The $\vmax$ values are taken from the halo catalog of a 160
Mpc/h simulation complete down to $\vmax \approx 90$km/s.  In order to
extrapolate this to lower circular velocities, we calculate a
power-law fit to the low end of the $dn/d\vmax$ function.  The
resulting correspondence is shown in Figure \ref{fig:abundance} for
the $r-$, $g-$, and $V-$bands (red, green, and black curves).  The $V$
band magnitudes are calculated using the transformation $V = g -
0.55(g-r) - 0.03$ from \cite{Smith02}.  This method implicitly assumes
that all galaxies have average color.  Since the data from
\cite{Blanton05} is not deep enough to map onto the dwarf galaxy
distribution, we use a power law to extrapolate the $M_V(\vmax)$
relation to lower magnitudes.  For the $V-$band, we get 
\be M_V - 5\log (h) = 18.2 -
2.5\log \left[\left({\vmax\over 1 \textrm{km/s}}\right)^{7.1}\right].
\label{eq:abundance}
\ee
When selecting the appropriate $\vmax$ for assigning a luminosity, we
follow the method of \cite{Conroy06} and choose the peak $\vmax$
over the trajectory of the subhalo for subhalos that eventually cross the $10^5\K$ post-reionization star forming threshold.  For subhalos that never reach this threshold,
we use the value of  $\vmax$ at $\zreion$. In both cases, this then corresponds roughly to the mass the halo had at the redshift they stopped rapidly forming stars.

The appeal of this method is that we are able to ignore much of the
poorly understood (and poorly simulated) physics of galaxy formation
using a statistical method that has been shown to, on average,
reproduce a wide variety of observable properties for more massive
galaxies \citep{Conroy06, Conroy08}, as well as some properties
of dwarf galaxies down to $v_{\rm max} \sim 50 km/s$
\citep{Blanton08}.  It is still unclear how this method will fare at
lower masses; it must break down for small halos once they no longer
host one galaxy on average.  If this transition is sharp, however, it
may be a reasonable approximation for most of the mass range where
halos host galaxies. 

As a second approach for assigning magnitudes, we use a toy model to
predict the star formation rate and stellar mass of a satellite
combined with the stellar population synthesis (SPS) code of
\cite{Bruzual03}\footnote{http://www.cida.ve/~bruzual/bc2003}.  
Here, we again assume that
star formation begins when the satellite first crosses the mass
threshold, $\mthresh$, and ends at the reionization time, $\zreion$.
During this period, the star formation rate is set by the
dark matter mass of the subhalo,
\be
SFR = \left\{ \begin{array}{ll}
\epsilon \left( f_{coldgas}{M_{DM} \over 1~\msol}\right)^\alpha &
\textrm{if } M_{DM} > \mthresh, ~z > \zreion \\
0 & \mbox{otherwise}
\end{array}
\right.
\label{eq:sfr}
\ee
where $f_{coldgas}$ is the fraction of cold gas in the halo, and
$\alpha$ and $\epsilon$ are free parameters.  This is similar to
model 1B of \cite{Koposov09}, with a couple of key differences.
First, we impose a hard truncation of star formation at the epoch of
reionization, something they only consider using their model where
stellar mass is a constant fraction of dark matter mass.  Second, they
treat stellar mass as being proportional $M_{DM}^{\alpha}$ at some
epoch, while we take the total stellar mass to be proportional to the
integral of this quantity, as motivated by observations
\citep[e.g.,][]{Juneau05, Noeske07, Zheng07, Conroy08}.
For our model, we hold $f_{coldgas}$ constant and set it equal to the
universal baryon fraction during the
period of active star formation.  Our two
free parameters $\epsilon$ and $\alpha$, set the efficiency and
scaling of the star formation.  We keep $\epsilon$ as a free parameter
that is used to match the luminosity function, but fix $\alpha = 2$,
extrapolated from higher mass galaxies at both low and high
redshifts \citep[Figure 8 of][]{Conroy08, Drory08}.
The
implicit assumption of this model is that subhalos contain a gas
fraction equal to the universal baryon fraction, and that this gas
exists in one of two phases.  At early times the gas in low mass halos
has a temperature less than $T = 8 \times 10^3\K$ and therefore cannot
cool by atomic hydrogen cooling.  
However, once a halo reaches
$\mthresh$, cooling becomes effective and all gas rapidly enters a
cold phase where it is able to form stars.  Reionization, however,
rapidly heats all gas in the subhalo to a few times $10^4\K$,
quenching star formation.  The more massive halos eventually shock
heat to such a virial temperature, which allows the ionized gas to
cool again and resume star formation.  Only 7 subhalos in the \VL
~simulation ever reach such a mass.  Regardless of mass, all
star formation is ended once a satellite accretes onto the larger host
halo, as we assume that this process causes all gas to be stripped from
the subhalo.  Most of the $z = 0$ subhalos in \VL ~were
accreted in the range $z = 0-6$.  This model does ignore a
large number of physical processes, such as recombinations and
feedback, but much of this can likely be accounted for by
appropriately setting the constants $\epsilon$ and $\alpha$.  In
Section \S5 we discuss how this model compares with previous studies.  
Satellite magnitudes for this model are determined using the stellar
population synthesis code of \cite{Bruzual03}.  For this model, we
treat the star formation as a series of bursts, all with the IMF of
\cite{Chabrier03} and metallicity $Z = 0.0004Z_{\odot}$, broadly
consistent with the expected level of enrichment from the earliest, most
massive stars that pre-enrich the halo gas \citep{Abel02,Wise08}.  
Using these magnitudes, we tune the $\epsilon$ parameter of equation
\ref{eq:sfr} so that our model most accurately reproduces the observed
luminosity function of local satellite galaxies (see below).  This
results in typical values for $\epsilon \sim 10\msol/$yr.

\begin{figure}
  \plotone{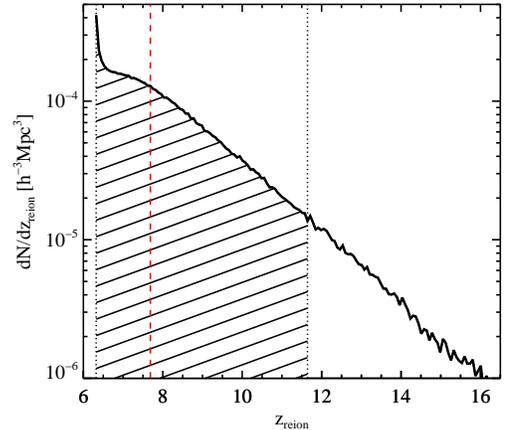} \figcaption{ Distribution of
    reionization epochs for Milky Way mass halos, $M_{FoF} = 1.6 - 2
    \times 10^{12}\hinv\msol$.  The dotted lines and hatched region
    indicate the 95\% limits for the distribution.  See
    \citet{Alvarez08} for more details on the simulations used to
    generate this distribution. The dashed red line indicates the
median value $\zreion = 7.7$.  
\label{fig:zreion}
}
\end{figure}

As a direct comparison, the resulting average $M_V(\vmax)$ relation from these two methods do not differ significantly.  When considering the populations as a whole, the SPS method results in a weaker magnitude trend with $\vmax$ and slightly reduced magnitude range.  On a galaxy by galaxy basis, however, this is appreciable scatter, with galaxy magnitudes differing by as much as 1.5 dex.  


\section{OBSERVATIONAL SAMPLE}

In the next section we present a number of comparisons with the
observed properties the Milky Way satellites.  We take as our data set
the 11 classical and 12 newly-identified SDSS dwarf galaxies within
the fiducial milky way virial radius of $417$kpc as
presented in Table I of \cite{Tollerud08}.  
However, we exclude Bo\"otes II from our comparisons of the maximum
circular velocity, $\vmax$, because there are only poor constraints.  
When using the SDSS dwarfs, we must account for both the sky coverage
of SDSS and the detection efficiency for each galaxy
\citep{Koposov08}.  While it is straight-forward to adjust the
observational data to account for sky coverage, a more subtle issue
arises with the depth of SDSS.  Because the survey is magnitude
limited, SDSS is only complete in searching for satellite galaxies of
absolute magnitude $M_v$ out to some radius $R_{comp}$.  This
completeness depth is roughly independent of surface brightness and
the relation is given in \cite{Koposov08} as \be R_{comp}(M_v) =
\left( {3 \over 4 \pi f_{DR5}} \right)^{1/3} 10^{(-aM_v - b)/3}
\rm{Mpc,}
\label{eq:rcomp}
\ee
where $f_{DR5}$ = 0.194 is the sky coverage fraction of SDSS, and a =
0.6 and b = 5.23. \cite{Tollerud08} also used the \VL
~simulation to calculate a correction due to this effect and use it to
correct the luminosity function.  However, this correction is unable
to account for the radial or $\vmax$ distributions.  Consequently,
whenever we compare our model predictions with observations, we
present the sky coverage-corrected distribution of observed satellites
and a sub-sample of the simulated galaxies that is cut by the
magnitude limit of equation \ref{eq:rcomp}.  

In addition to survey depth, the observational sample of Milky Way dwarfs is also limited by a surface brightness detectability constraint.  This is discussed in detail in \cite{Koposov08} who parameterize the detection efficiency of a satellite according to
\be
\epsilon(M_v,\mu) = G\left({M_v - M_{v,lim} \over \sigma_M}\right)
G\left({\mu_v - \mu_{v,lim} \over \sigma_{\mu}}\right),
\label{eq:mag_mu_lim}
\ee
where $G(x) = 0.5{\rm erfc}(x/\sqrt{2})$ and $M_{v,lim}$, $\mu_{lim}$, $\sigma_M$, and $\sigma_{\mu}$ describe the magnitude and surface brightness limits of observations.  Using their parameters for fitting current observations, we calculate the detection efficiency of our model galaxies using the conservative assumption that the light of a satellite is uniformly distributed out to the tidal radius of its host dark matter subhalo.  

Finally, we must account for the observational uncertainties in measurements of the magnitudes of the dimmest dwarf galaxies.  Effectively, the small number of stars used to determine the magnitude of a faint dwarf galaxy introduces shot-noise effects, creating additional uncertainty on the measurement.  This is studied in detail in \cite{Martin08}, who report magnitude uncertainties in their Table 1.  We include these uncertainties when making comparisons between our modeled luminosity function and observations.  

\section{RESULTS}

\subsection{Distribution of reionization epochs}

First, we first investigate the distribution of reionization times for
Milky-Way size halos in Figure \ref{fig:zreion}.  The solid line of
the Figure shows the distribution of reionization epochs for the
$\sim$500,000 identified halos with mass $M_{FoF} = 1.6 - 2 \times
10^{12}\hinv\msol$ from our Gpc simulation. From this plot, we see a
wide distribution of reionization epochs, peaked around $z = 6$ with a
tail extending beyond $z = 16$, fit by an exponential, $dN/d\zreion
\propto e^{-0.6\zreion}$.  Ninety-five percent of the halos are
reionized in the
range $\zreion = 6.3 - 11.6$ with a median redshift of 7.7. This
distribution is also quite consistent with similar predictions from
\cite{Weinmann07}. Percolation of reionization happens at $z = 6.3$
and the universe rapidly becomes fully ionized, which is responsible
for the sharp cutoff in the distribution.  This broad distribution
indicates that a precise constraint on the globally-averaged
reionization epoch, for example as measured from the optical depth to
the cosmic microwave background, does not give precise constraints on
the reionization history of our local galaxy and its progenitors.  As
we will show, determining where the Milky Way itself sits in this
distribution may be important for understanding its satellite
population.  The reionization epoch of a given halo is almost
certainly correlated with other properties, including the large scale
bias and the detailed formation history of the halo, and there may be
observational clues beyond those presented here, to where the Milky
Way lies in the distribution. We postpone an investigation of these
issues to future work.


\subsection{Distribution of satellite populations}

\begin{figure}
\plotone{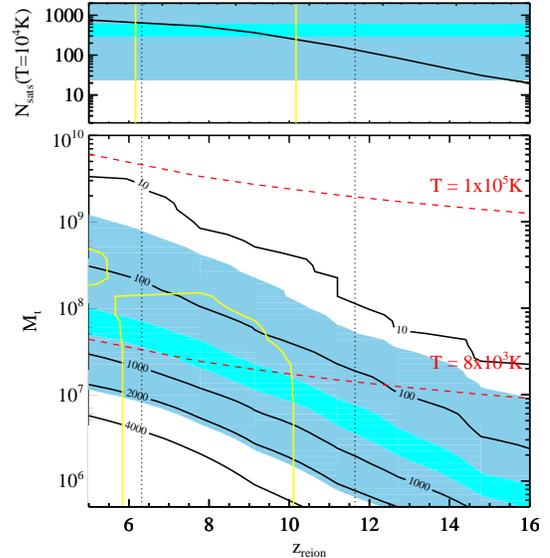}
\figcaption{
{\it Bottom: } The number of surviving satellite galaxies in the \VL simulation assuming that subhalos had to grow above a threshold
mass $\mthresh$ by $\zreion$ in order to host a galaxy.  The dashed
vertical lines show the 95\% confidence intervals for the reionization
epoch of Milky Way mass halos.  The dashed red lines denote constant
virial temperatures, where $8 \times 10^3K$ is the expected temperature for
allowing star formation through HI cooling.  The the light and dark
shaded regions show the observational constraints on the both the
maximal and most likely distribution in the number of satellite
galaxies from \cite{Tollerud08}.  The yellow contours denote the 95\%
confidence levels for constraining $\zreion$ and $\mthresh$ by
matching with the $\vmax$ function of Figure \ref{fig:vmax}.  {\it
Top: } The number of satellite galaxies as a function of $\zreion$
assuming that $\mthresh$ follows the $T = 8 \times 10^3K$ curve.  The
dotted lines show the 95\% distribution for $\zreion$ for Milky Way
mass halos, and the yellow lines show the 95\% confidence range in
$\zreion$ at this $\mthresh$.  Again, the light and dark cyan regions
show full and most likely range for the number of Milky Way satellite
galaxies.
\label{fig:nsats}
}
\end{figure}

Using the distribution of $\zreion$ from Figure \ref{fig:zreion} as a guide, we show in Figure \ref{fig:nsats}
the number of subhalos hosting satellites galaxies, $\nsats$, in the
\VL ~simulation for our model.  The contours of the lower panel
show the full variation in $\nsats$ as a function of both $\zreion$
and $\mthresh$.  The dashed lines denote constant virial temperatures.
The thick lower line represents $\tvir = 8\times 10^3\K$, the
temperature where HI cooling is expected to be effective for gas that
has not been photoionized \citep{Wise08}.  The upper line is $10^5\K$,
the mass where the halo has shock heated to the point where
photoionized gas can cool \citep[e.g.,][]{Haiman06}.  The top panel shows
$\nsats$ as a function of $\zreion$ when $\mthresh$ is set such that
star formation begins once the subhalo reaches a virial temperature
$T_{vir} = 8 \times 10^3K$.  This line asymptotes to $\nsats = 7$
because we allow star formation to resume after reionization for halos
that shock heat to temperature $\tvir = 10^5\K$.  Virial temperatures
are calculated using the relation 
\bee 
M = 10^6~\msol\left[\left({\tvir \over 1800. \K}\right) \left({21\over
      1+z}\right) \left({1.22 \over \mu}\right)\right]^{3/2}
\times \nonumber \\
\left({0.3 \over \Omega_{M,0}}\right)^{1/2} \left({0.7 \over
    h}\right), 
\eee 
where $\mu = 0.57$ for $\tvir < 1.5\times 10^4 \K$
and 1.22 for $\tvir \ge 1.5 \times 10^4 \K$ \citep{Haiman06}.  
The model referred to as "all galaxies" in this plot indicates all subhalos capable of hosting a galaxy. 
The galaxies are not defined in terms of a magnitude limit, but weather
they reached a mass scale (left as a variable in this model) above
which star formation was allowed to proceed.  As a consequence, many
such galaxies are expected to be extremely low luminosity, well below
current detection limits.  Additionally, the model is simply a
function of the subhalo mass at times prior to accretion and does not
track effects such as tidal disruption.

The most
striking feature of this plot is the size of the variation in
$\nsats$, which can differ by roughly an order of magnitude at fixed
$T(\mthresh)$ as $\zreion$ is varied within the 95\% distribution or
as $\mthresh$ changes from $\tvir = 10^3 - 10^4\K$.  For a constant
$\mthresh$, the number of satellites has a roughly exponential
dependence on $\zreion$, $\nsats \propto e^{-0.75\zreion}$.  Again, there is
an imposed minimum value of $\nsats = 7$ for any value of $\zreion$, 
set by the assumption that photoionized gas can cool and form stars when a
halo shock heats to $\tvir = 10^5\K$; seven of the \VL
~subhalos reach this temperature at some point in their histories.

The shaded region of Figure \ref{fig:nsats} shows the observational
constraints on the number of satellite galaxies from the work of
\cite{Tollerud08}.  The lighter regions shows the extreme case of this
analysis, where the Milky Way has between 23 (currently observed) and
2261 satellites galaxies, while the darker region shows what their
work considers the more likely prediction of 300--600 satellites.  
When all this data is combined, the consistency is rather surprising.
The region spanned by the upper and lower 95th percentile for
$\zreion$ of Milky Way mass halos, as well as the $\mthresh \approx
10^6 \msol\hinv$ from \cite{Abel02} and the $\tvir = 8 \times 10^3\K$
falls entirely within the constraints of \cite{Tollerud08}.  It is
important to note that these results rely on simulations using
completely different sets of physics.  The limits on $\mthresh$ are
set from hydrodynamical simulations as discussed in
\S\ref{sec:galform}, while the $\zreion$ and $\nsats$ limits come
from N-body simulations that model only collisionless physics.  

One caveat that must be kept in mind when interpreting Figure
\ref{fig:nsats}, is that we have assumed the particular subhalo
population of \VL ~is representative of a typical Milky Way mass halo.
While only a handful of such
ultra-high resolution simulations have been conducted, it is already
apparent that there is a wide distribution in the number of subhalos
in halos of similar mass.  In particular, currently the three most
well resolved halos \citep{Diemand08, Stadel08, Springel08} contain a
factor of 1.5$-$2 more subhalos than \VL ~at a fixed mass threshold, and
it is estimated that \VL ~is among the 15\% of objects with similar
mass that have so few subhalos \citep{Ishiyama08}.  If this is the
case, the number of satellites predicted in Figure \ref{fig:nsats} is
potentially a factor of two too low for a typical Milky Way mass
halo. However, it is unknown exactly where in the relatively wide
distribution the Milky Way lies, particularly since the number of
subhalos has been shown to correlate strongly with halo concentration
and formation history \citep{Zentner05}.  In the remainder of this
work, we assume that the subhalos in \VL ~are representative of the
Milky Way, but this distribution, and the possible bias, should be
kept in mind when detailed numerical results are given.

\subsection{Luminosity Function}

\begin{figure}
\plotone{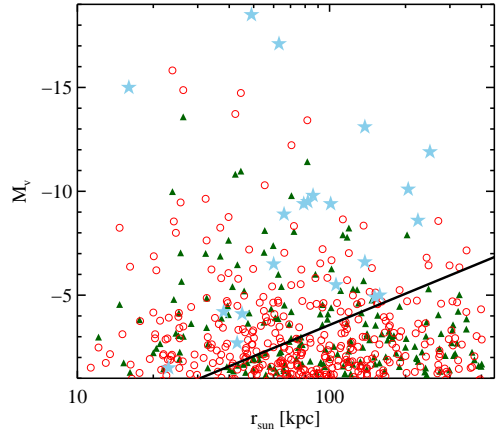}
\figcaption{
The distributions of satellite galaxy magnitudes as a function of
distance from the sun in a model with $\zreion = 9$ and $\mthresh = 3
\times 10^7\hinv\msol$.  The open red circles show magnitudes assigned
using the abundance matching method (equation \ref{eq:abundance}) and
the filled green triangles have magnitudes assigned using the Bruzual
and Charlot stellar population synthesis (SPS) code.  The cyan stars show
the distribution of the observed Milky Way satellites.  The solid line
shows the completeness depth of the SDSS survey as given by equation
\ref{eq:rcomp}.  
\label{fig:rcomp}
}
\end{figure}

While Figure \ref{fig:nsats} shows that the total number of subhalos
hosting satellite galaxies may be strongly dependent on the time of
reionization, it is necessary to understand the properties of these
affected halos, i.e., are they all low mass objects that we expect to
host low-luminosity galaxies, or do they fill a larger range in
satellite parameter space?  In order to quantify the expected impact
on observations, we must first impose the relevant observational cuts
on our satellite distribution.  For each subhalo, we calculate
$r_{sun}$, the distance from a point 8kpc from the center of \VL.
Figure \ref{fig:rcomp} shows this distribution as a function of
magnitude for the model $\zreion = 8$, $\mthresh = 3 \times
10^7\hinv\msol$.  The open red circles show magnitudes calculated
using the abundance matching method (equation \ref{eq:abundance}), and
the filled green triangles use the population synthesis model. We then
impose the cut defined by equation \ref{eq:rcomp} above, shown as the
black line.  Because we expect this subset to best match the
observational sample, this cut is imposed for all subsequent
comparisons.  While only affecting about 20\% of our satellites,
objects as bright as $M_V = -7$ are cut.  The distributions of the
Milky Way dwarfs are overplotted for reference.  We also calculate the detection efficiency for each satellite galaxy based on its surface brightness according to equation \ref{eq:mag_mu_lim}.  All galaxies passing out magnitude cut have $\epsilon \approx 1$, so we do not make any additional cuts based on estimated surface brightness.

Again, because the magnitudes set by the abundance matching method are
not directly impacted by $\zreion$ and $\mthresh$, the distribution of
objects in $M_V-r_{sun}$ space in not strongly impacted as these
parameters are varied.  In particular, adjusting these parameters
only results in the presence or absence of objects with low $M_V$ as
low mass subhalos gain or lose the ability to host satellite
galaxies.  Individual objects will, however, have a significant
dependence on magnitude in the SPS model because adjusting these
parameters impacts how long star formation is allowed to proceed for,
impacting the amount of mass that can be converted into stars.  In
addition to forming new satellite galaxies, pushing $\zreion$ to later
epochs also causes the existing satellites to brighten.  

\begin{figure}
\plotone{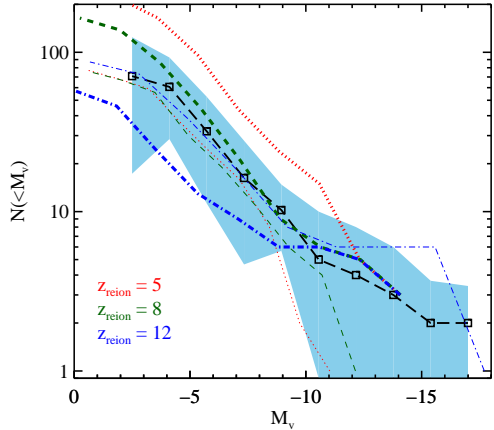}
\figcaption{
Luminosity functions for observations and model predictions.  The
long-dashed line shows the observed Milky Way satellite luminosity
function corrected for sky coverage and depth effects, while the cyan
swath represents the statistical error.  The red
dotted, green dashed, and blue dot-dashed lines represent reionization
models of varying $\zreion = 5, 8$, and 12, respectively.
$\mthresh$ is set using the virial temperature, $\tvir(\mthresh) = 8
\times 10^3\K$.  The thicker set of lines shows predicted luminosity
functions using an extrapolated abundance matching method to assign
luminosities to the galaxies.  The thinner set of lines use a stellar
population synthesis model to predict the luminosities.  
\label{fig:lf}
}
\end{figure}

Figure \ref{fig:lf}  compares the luminosity functions from our model
with observations.  The thicker lines show magnitudes set by the
abundance matching method, and the thinner lines by the SPS model.  
For this plot, we have fixed $\mthresh$ to be set by the $\tvir(\mthresh) =
8\times 10^3\K$ relation and varied $\zreion = 5, 8, {\rm and ~}
12$ (red dotted, green dashed, and blue dot-dashed lines).  
The black long-dashed line with points shows the observed
luminosity function, while the cyan region shows the Poisson
errors about this distribution.  

Overall, both the abundance matching and SPS methods reproduce the
observed luminosity reasonably well.  The level of agreement for the
abundance matching method in particular is rather impressive since the
method 1) ignores all baryon physics such as star formation, 2)
assumes all galaxies have average color, and 3) extrapolates number
densities down to regimes where the method has not been tested and
where small-scale processes may produce a significant
amount of scatter in the $\vmax-M_V$ relation.  Because the primary
effect of changing $\zreion$ is to change the total number of
satellite galaxies, this serves to change the overall normalization of
the luminosity function while retaining the slope.  Although not shown
here, changing $\mthresh$ for a fixed $\zreion$ has a similar effect.

The high level of agreement between our SPS model and
observations comes from tuning the efficiency
parameter, $\epsilon$, to match observations independently for all
values of $\zreion$.  Thus, the three curves in Figure \ref{fig:lf}
have values $\epsilon = 0.1, 10, \textrm{ and } 300 \msol/{\rm yr}$,
which were selected for no physical reason other than to match the
luminosity function.   It is, however, interesting to note that the
faint-end slope of this model almost perfectly matches the
observations.  The under-prediction of luminosities at the bright end can
potentially be explained by residual, ongoing star formation.  At
least one of the classical Milky Way Dwarfs, Leo I, shows significant
evidence -- while Fornax shows slight evidence -- for recent star
formation after the epoch of accretion \citep{Mateo98, Mateo08}.  If
we were to allow some such process in our 4 brightest objects the
model will likely fit the observations significantly better.

\begin{figure}
\plotone{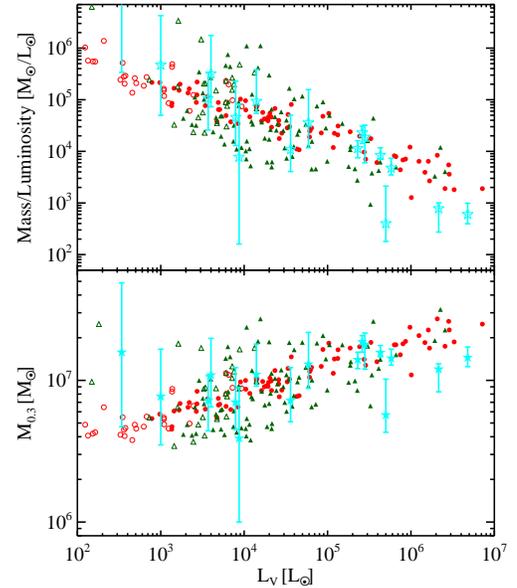}
\figcaption{
{\it Top Panel: } The ratio of subhalo mass at the time of accretion
to luminosity for our abundance matching model (red circles) with
$\zreion = 8$.  Filled circle represent halos that are within the
SDSS completeness radius, equation \ref{eq:rcomp}, while open circles
are outside this distance.  Cyan stars represent Milky Way satellites.
In order to make this comparison, we used the values for mass within
300 kpc at the present epoch, $M_{0.3}$, published in
\cite{Strigari08} and converted them to subhalo masses using their
published relation for the average $M_{DM}(M_{0.3})$ calibrated to
N-body simulations.  
{\it Bottom Panel: } The relation between $M_{0.3}$ and luminosity for
our abundance matching model and observations.  Here, we have
converted the masses of the \VL ~subhalo to $M_{0.3}$ values
using the formula provided by \cite{Strigari08}.  
\label{fig:ML}
}
\end{figure}

Finally, we consider the ration of mass to luminosity in the top panel
of Figure \ref{fig:ML}.  Here, we plot the mass to light ratio of the
\VL ~subhalos using their virial masses at time of accretion and
luminosities assigned from the abundance matching method for $\zreion
= 8$ as red circles.  Solid circles represent objects
within the SDSS magnitude limit, $r_{sun} < R_{comp}(M_V)$, where
$R_{comp}$ is given by equation \ref{eq:rcomp}.  Open circles are
satellites outside this limit.  This model naturally
reproduces a wide range of $M/L$ rations, spanning from $10^3$ to
$10^6$, with a clear trend with luminosity, $M/L \propto L^{-0.52}$.
Green triangles represent luminosities calculated using our SPS
method.  This method reproduces a similar trend with luminosity but
with a larger dispersion.  
We have
included measurements from the Milky Way as the
cyan stars.  In order to model the masses of these objects, we took
the $M_{0.3}$ values for the mass within 300kpc published in
\cite{Strigari08} and converted those to subhalo masses using their
published relation $M_{0.3} = 10^7\msol(M_{vir}/10^9\msol)^{0.35}$,
where $M_{vir}$ is the virial mass of the subhalo at the time of
accretion.  We must caution that this relation is expected to be
dependent on cosmology and ignores all scatter.   The observations are
remarkably well matched by our model with excellent agreement for all
but the most most luminous galaxies.  
In the lower panel of Figure \ref{fig:ML} we consider this data in a
different way by plotting $M_{0.3}$ as a function of luminosity.
Here, we take the data directly from \cite{Strigari08} and convert the
\VL ~subhalo masses to $M_{0.3}$ using the above formula.  While the
numbers are in general agreement, the abundance matching model (red
circles) shows a clear trend with luminosity, $M_{0.3} \propto
L_V^{0.17}$, as opposed to the observations (cyan stars) which
indicate more of a common mass scale.    
The trend in our model results directly from
the abundance matching method, which assigns luminosities to subhalos
satellite based on $\vmax$ at the time of accretion.  The addition
of scatter into either the $L_V(\vmax)$ or $M_{0.3}(M_{vir})$
relations, which we expect at these low mass scales,
can help to flatten this trend slightly and bring it more in line with
observations.  
The SPS model (green triangles) also produces a similar trend with
luminosity albeit with a significantly larger scatter, making the
slope of the $M_{0.3}(L_V)$ relation consistent with zero.  
Previous studies of simulations
have observed a similar trend \citep[e.g.,][]{Maccio08,
Li08, Koposov09} using semi-analytic modeling of galaxies and/or subhalo
distributions.  

\subsection{Circular Velocity and Radial Distributions}

\begin{figure}
\plotone{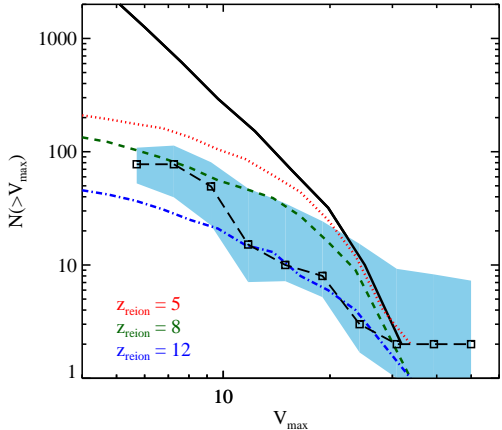}
\figcaption{
The peak circular velocity functions for subhalos
hosting satellite galaxies.  The solid line shows the velocity
function for all \VL ~subhalos. The red dotted, green dashed,
and blue dash-dotted lines show the distribution for subhalos hosting
satellite galaxies for $\zreion = 5, 8$, and 12 with
$\tvir(\mthresh) = 3\times 10^8\K$, as in Figure \ref{fig:lf}.  The
long dashed black line shows observed distribution for Milky Way
satellites, corrected for sky coverage and detection efficiency.  The
cyan bands show combined Monte Carlo and statistical errors.  
\label{fig:vmax}
}
\end{figure}

We next consider the mass distribution of the satellite galaxies
hosting halos.  
Figure \ref{fig:vmax} shows the changes in the $\vmax$ distribution
for satellites as $\zreion$ is varied, given
a threshold mass of $\tvir(\mthresh) = 8 \times 10^3 \K$ as in Figure
\ref{fig:lf}.  Here, the solid black line shows the distribution for
all \VL ~subhalos, while the red dotted, green dashed, and blue
dot-dashed show the distributions from our model for three values of
$\zreion = $ 5, 8, and 12.  Because the abundance matching
method was more successful than the SPS model in reproducing the
observed luminosity function without the need to tune any parameters, we
only include satellites that pass the radial cut of equation
\ref{eq:rcomp} using the abundance matching criteria.  
As can be seen, an earlier
$\zreion$ suppresses the distribution of subhalos with all values of
$\vmax$, although the effect is more pronounced for low mass halos.
Still, this suppression is present even for $\vmax \gsim 20$km/s,
where most of the classical dwarfs live.  This indicates that
$\zreion$ can effect satellite galaxies of all masses and luminosities.

The long-dashed black line with data points again represents the
observations.  The $\vmax$ values for the satellites, including
errors, were calculated using the method of \cite{Strigari07c,
  Strigari07d} using kinematic data taken from the literature
\citep{Walker07, Simon07}.  The line was calculated using the 22
observed satellites and correcting them for SDSS sky coverage and
detection efficiency \citep{Koposov08}.  The cyan region denotes
errors on this curve and were calculated using a Monte Carlo approach.
In this approach, published errors are used where possible; where no
robust errors are published, the average error distribution is mapped onto
the remaining SDSS dwarfs.  While this process does not produce
uncorrelated error bars, it should be significantly more robust than
simply using statistical uncertainties. We should also note that,
because the reconstruction of $\vmax$ from observations gives a very
strong lower bound but only a weak upper bound, the errors on the
lowest points are probably underestimated because very few satellites
will scatter into this bin as we create a Monte Carlo representation
of the distribution.  These systematic errors were combined with
statistical errors assuming a Poisson distribution.

\begin{figure}
  \plotone{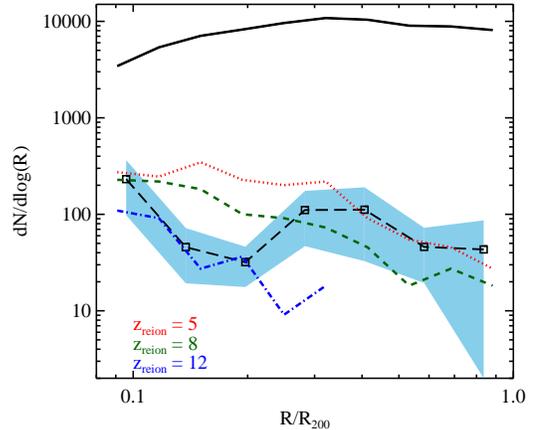} \figcaption{ The radial
    distribution subhalos hosting satellite galaxies in \VL.  The
    lines represent the same populations as in Figure \ref{fig:vmax}.
\label{fig:rhalo}
}
\end{figure}

Figure \ref{fig:rhalo} further explores the impact of varying
$\zreion$ on the properties of the satellites and the subhalo hosts by
considering the radial distribution within the halo.  The lines represent the same
populations as in Figure \ref{fig:vmax}.  Here, the errors on the
observations are purely Poisson.  Again, there is strong trend for an
early reionization epoch to suppress the abundance of satellite
galaxies at all radii, but, in part because the data is rather noisy,
it remains easy to match the model to the observations.

Finally, using our magnitude-limited sample of satellites from the
abundance matching method, we can attempt to put constraints on
$\zreion$ and $\mthresh$.  We calculate the $\chi^2$ errors between
the $\vmax$ distribution for observations and our models and plot the
95\% confidence levels as the yellow contour in Figure
\ref{fig:nsats}.  We use the $\vmax$ distribution to generate this
constraint because it is less sensitive to the modeling of the
magnitudes than the luminosity function is, with the modeling only
coming into play when making the SDSS magnitude-limit cut.  Because
the Monte Carlo method underestimates the error on the lowest $\vmax$
point in Figure \ref{fig:vmax}, we exclude it from our constraints.
The general insensitivity of our constraints to the value of $\mthresh$ comes from our use of $\vmax(\zreion)$ in setting the magnitude of a galaxy.  While lowering the threshold for star formation from $\sim 10^8$ to $\sim 10^9\hinv\msol$ significantly increases the number of subhalos hosting galaxies, most of these objects are quite dim, with $M_V \gsim -4$, and are therefore cut from our observable sample (see Figure \ref{fig:rcomp}).  
If we impose the condition that galaxy
formation begins when a halo reaches virial temperature $\tvir(\mthresh) = 3
\times 10^8\K$ we can constrain the time of reionization to $\zreion =
8^{+3}_{-2}$, with $540^{+100}_{-340}$ satellites.  Note that
these errors assume that the dark matter substructure of the Milky Way
is identical to that of the \VL ~halo; if the Milky Way is more typical
for its mass it may have a larger total number of observable
satellites.  A further success of this model is that, when we
constrain $\zreion$ using the $\vmax$ function, we naturally reproduce
both the observed luminosity function and radial distribution, as
shown by the green dashed lines in Figures \ref{fig:lf} and
\ref{fig:rhalo}.  We should, however, caution again that much of this
depends on the particular realization that is the \VL ~halo, and that
more statistics will be necessary for a more robust prediction.

\section{COMPARISON WITH PREVIOUS WORK}

Our result that the number of satellite galaxies is strongly dependent
on the redshift of reionization is at odds with some previous studies,
including the work of \cite{Somerville02}, \cite{Kravtsov04b}, and \cite{Orban08}, 
although these studies primarily focused on the classical satellite
galaxies in the Milky Way.  The primary difference in our models rests
on the assumption of how reionization effects the presence of cold
gas.  These previous studies have used the model of \cite{Gnedin00} to
calculate the amount of cold gas in a halo of mass $M$.  They
calculate a filtering mass, $M_f(z)$, the mass of a halo that looses
half its baryons compared to the universal baryon fraction in the
presence of a photoionizing UV background.  This mass is related to
the baryon fraction of a halo of mass $M$ via the relation \be f_{\rm
  gas}(M,z) = {f_{\rm baryon} \over [1+0.26M_f(z)/M]^3},
\label{eq:fgas}
\ee
where the details for calculating the filtering mass, $M_f$, are given
in \cite{Gnedin00} and Appendix B of \cite{Kravtsov04b}.  

The crucial point in using the values of $f_{\rm gas}$ from
\cite{Gnedin00} lies in how these fractions are related to the amount
of {\em cold} gas available for star formation.  There are two
bracketing possibilities: either the gas is spread out over the entire
halo and is hot, in which case the star formation rate is zero, or,
there is a small clump of cold gas in the center, with a mass given by
$f_{\rm gas}M_{\rm halo}$, that can form stars.  Our interpretation, in
which the gas is assumed to be hot for $f_{\rm gas}\ll 1$, is
consistent with the first possibility, while these previous studies
have implicitly assumed that all the gas is cold.  

There are several reasons to favor the hot gas scenario.  Most importantly,
\citet{Gnedin00} did not distinguish between hot and cold gas when
calculating the gas fraction.  The assumption that the gas is cold is
only valid for halos that had collapsed before the reionization epoch,
and were then subject to photo evaporation due to the UV background.
Such cold gas would only survive for one photo evaporation time, which,
according to the numerical simulations of \citet{Iliev05}, is
likely to be less than 500 Myr, corresponding to $\Delta
z<2$ at $z<6$.  After such photo evaporation, the halo could
only accrete hot gas, which would be unlikely to form stars given its low
density and long cooling time.  Moreover, due to the exponential growth
of the abundance of halos with masses $\sim 10^8 M_\odot$ during reionization,
it is likely that most halos present after reionization were only just
collapsing, and thus did not have any cold gas capable of forming
stars in the first place.  We therefore interpret the gas fractions
$f_{\rm gas}\ll 1$ reported by \citet{Gnedin00} after reionization as
corresponding to hot gas when $f_{\rm gas} \ll \Omega_b/\Omega_m$. 

Another related caveat in using $f_{\rm gas}$ to model the dependence
of star formation history on reionization is that the value of $f_{\rm
  gas}$ reported by \citet{Gnedin00} is a {\em mean} value, for all
halos in the box, regardless of whether they are in ionized regions or
not.  In reality, those halos that were reionized earlier have lower
gas mass fractions than those that were reionized later.  Using the
average value of $f_{\rm gas}$ for all halos underestimates the
sharpness of the transition for halos that were ionized at a given
time.  Because our reionization redshift is defined for a {\em given}
halo (as opposed to a universal time for the universe), we expect the
transition in gas fraction to be much sharper than that given by
averaging over all halos, each with differing values of $z_{\rm
  reion}$, as was done by \citet{Gnedin00}.  

\begin{figure}
\plotone{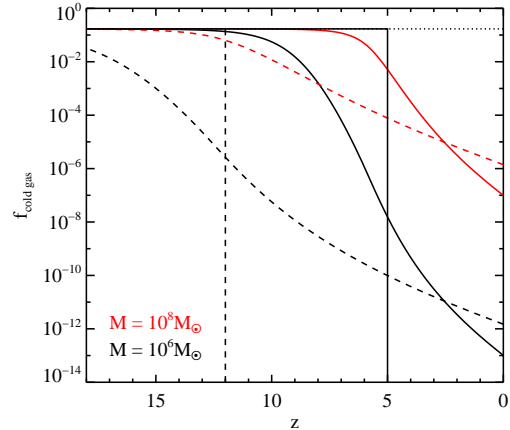}
\figcaption{
The fraction of mass in the form of cold gas as a function of time.
The solid  and dotted lines are calculated using {\sl universal}
reionization epochs $z = 5$ and 12, respectively.  The lower
(black) and upper (red) pairs of curves show the predictions from
\cite{Gnedin00} for halos of mass $M_{DM} = 10^6 \hinv \msol$ and
$10^8 \hinv \msol$, respectively.  The vertical lines represent the
model used in this work, where the reionization time of the subhalo is
$\zreion = 6.5, 11.5$ for the solid and dashed lines.  The horizontal
dotted line shows the universal baryon fraction.  
\label{fig:fgas}
}
\end{figure}

The difference between these scenarios --- where $f_{gas}$ represents
the fraction of hot in a halo consistent with our abundance matching
and SPS models as opposed to the fraction of cold gas available for
star formation --- is illustrated by Figure \ref{fig:fgas}, which
shows the evolution of $f_{\rm gas}(M,z)$, the material available for star
formation, with redshift.  The upper and lower pairs of lines in the
figure represent halos with mass $M = 10^8 \hinv \msol$ and $10^6
\hinv \msol$.  The solid and dashed lines represent models where the
universe reionizes at different epochs, $z = 5$ and12,
respectively.  At high redshift, halos have a gas fraction equal to
the universal baryon fraction.  As time evolves, the $f_{\rm gas}$ of
the average halo decreases rather slowly, due to photoionization
heating as halos begin to be exposed to the UV background.  Changing
the redshift of reionization affects the rate of transition from the
cold to the hot phase, but the overall shape of the transition is
preserved.  As we emphasized above, because the average over all halos
was used to calculate the filtering mass, this is likely an accurate
description of how reionization effects the gas content of an {\sl
  average} halo of mass $M$ given a universal redshift epoch.  We
contrast this with the model used for this work, shown by the vertical
lines indicating a sharp cutoff in the presence of cold gas at the
time at which the halo reionizes is mass independent.  While we
acknowledge that we are ignoring effects such as atomic recombination
and additional cooling for high mass objects that should not quite
reduce the cold gas fraction to zero after reionization, we expect
that such a sharp transition from the cold to hot phase more
accurately describes the evolution of an individual halo.

\begin{figure}
\plotone{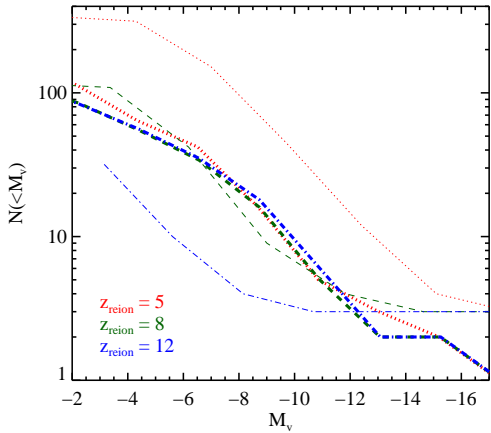}
\figcaption{
The luminosity functions from our SPS model for both the case of an
instantaneous gas heating (thin lines) and a more gradual turnoff as
given by equation \ref{eq:fgas} (thick lines).  Colors and line-styles
represent different epochs of reionization as in Figure \ref{fig:lf}
\label{fig:lf_gnedin}
}
\end{figure}

The difference in the resulting star formation histories in the two
scenarios discussed above is dramatic.  To show this explicitly, we
have re-run our 
population synthesis model using the more gradual 
star-formation squelching model of equation \ref{eq:fgas} instead of
an abrupt squelching for setting $f_{coldgas}$ in equation
\ref{eq:sfr}, although we still assume an abrupt end to star 
formation when the subhalo is accreted onto the host halo.  The result,
shown in Figure \ref{fig:lf_gnedin}, is that the luminosity function of the
satellite galaxies becomes largely independent of the reionization
epoch, in agreement with these previous studies.  Here, the thin
lines represent our model with
an instantaneous gas heating, while the thick lines show the gradual
turn off from equation \ref{eq:fgas}.  The red dotted, green dashed,
and blue dot-dashed show the effect of changing the reionization time,
$\zreion = 5, 8, 12$.  We should note that, unlike in Figure
\ref{fig:lf}, we did not tune the parameters $\alpha$ and $\epsilon$ of
equation \ref{eq:sfr} to reproduce the observational sample at all
values of $\zreion$, but just set the parameters to fit the $\zreion =
9.6$ model.  Additionally, this plot does not include any observational magnitude cuts, equation \ref{eq:rcomp}.  While the instantaneous-squelching model shows
dramatically different luminosity functions as $\zreion$ is changed,
the gradual cutoff model is remarkably stable.  This is because, even
though the rate at which gas heats varies with reionization time, the
gradual turnoff causes roughly the same amount of stellar mass to form
in an average halo regardless of the reionization time.  The average stellar
mass, $\langle\log(M_{stellar})\rangle$, changes by less than 5\% as $\zreion$
varies from 12 to 5.  For our rapidly
truncating model, however, the amount of gas converted into stars
clearly depends on the time of reionization.  
This raises an important point, in that while there is
some disagreement as to the impact the time of reionization has on the
satellite galaxy population of a halo, the rate at which star
formation is squelched can also have a significant impact on the
population. 

In order to understand why the luminosity function does not vary with
$\zreion$ in this model, it is necessary to look at the magnitudes of
individual galaxies.  The most massive subhalos in \VL ~are
larger than the filtering mass, $M_f$, and are therefore unaffected by
changes in the reionization epoch.  In order to understand why the dim
end of the luminosity function has such a weak dependence on the
reionization epoch, we note that, as pointed out in \cite{Koposov09},
halos in this model with a gradual turnoff of star formation create a
large number of stars after $\zreion$.  Additionally, as seen in
Figure \ref{fig:fgas}, there is a transition epoch around $z
= 3$ between suppression and enhancement of $f_{\rm gas}$ for early
and late reionization times.  At early epochs, an early $\zreion$
suppresses the amount of gas in a halo of a given mass relative to a
late reionization because of the additional energy input to the
system.  However, at late times, the early $\zreion$ actually causes
an enhancement relative to later reionization because the expansion of
the universe since reionization causes adiabatic cooling.  This
transition is roughly independent of mass.  Thus, there are two
regimes: Subhalos that accrete onto the main halo early, before $z =
3$, must get dimmer as $\zreion$ increases, because star
formation is suppressed at all epochs.  However, objects that
accrete recently pass through a phase where earlier reionization
enhances star formation, potentially allowing such satellites to brighten
with.  

\begin{figure}
\plotone{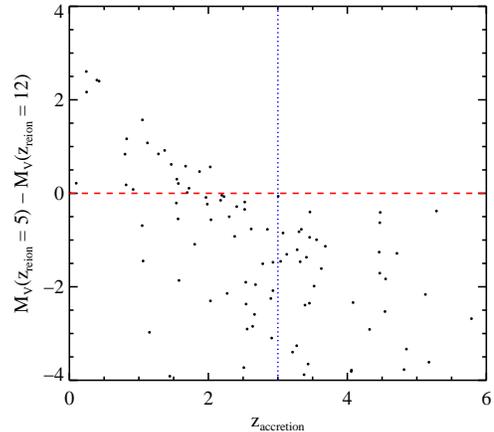}
\figcaption{
The change in magnitude for satellite galaxies between late ($z =
12$) and early ($z = 5$) $\zreion$ using the model motivated by
\cite{Gnedin00}, where the amount
of cold gas is given by equation \ref{eq:fgas}.  Galaxies that have
been accreted more recently are brighter for earlier $\zreion$ due to
the boost in star formation from adiabatic cooling.  The opposite
trend is present for galaxies that are accreted earlier.
\label{fig:Mv_zac}
}
\end{figure}

This trend between star formation and time is seen in Figure
\ref{fig:Mv_zac}, which plots $\Delta(M_V)$, the difference in
magnitude for individual satellite galaxies between early ($\zreion =
12$) and late ($\zreion = 5$) reionization, 
as a function of accretion time within the context of a gradual star
formation turnoff. Positive values represent satellites that brighten
with early reionization, while negative values represent objects that
get dimmer with early reionization.  There is a clear trend with
$z_{accretion}$:  Prior to $z = 3$, all objects have negative values
for $\Delta(M_V)$, as expected from Figure \ref{fig:fgas}, while
objects that accreted more recently have both positive an negative
values, with the most positive values (strongest brightening from
earlier reionization) coming from the most recently accreted
satellites.  Thus, while $\zreion$ does effect the magnitude of a
given satellite in the context of a model with a gradual star
formation turnoff, there are two competing effects that cause the
luminosity function to be unchanged.  This explains the differences between
our results and those of \cite{Somerville02} who
concluded that $\zreion$ has no effect on the satellite galaxy
luminosity function.  Hydrodynamical simulations also tend to disfavor
this scenario.  As shown in \cite{Abel99} and \cite{Sokasian02},
quasar activity around $z \sim 3-4$ would have resulted in an epoch of
HeII reionization, something not considered in the calculations of
\cite{Gnedin00}.  This would have resulted in additional heating of
the gas at these epochs, likely destroying both the enhanced
recent star formation activity due to early reionization and
independence of the satellite galaxy luminosity function on the epoch
of reionization.  

We also reach different conclusions than the recent work of
\cite{Koposov09} as to whether an abrupt or a gradual star formation
truncation more accurately matches observations.  In particular,
they note a bimodality in the luminosity function for an abrupt cutoff
to star formation.  Their model differs from ours, however, in the key
respect that they assume a halo converts a constant fraction of
its gas mass into stars, resulting in stellar mass $M_* \propto
M_{DM}$, where $M_{DM}$ is the dark matter mass at either
reionization or $z = 0$, depending on the mass of the halo.  This
results in a low-luminosity population of galaxies unable to form
stars after reionization, and a higher luminosity population that can
form stars.  If we adopt such a model and apply it to the \VL
~subhalos we reproduce a similar set of populations, with a population
of dim satellites in a very narrow luminosity range separated from a
brighter population made up of the 7 most massive subhalos that are
able to sustain ongoing star formation after the reionization epoch.
As \cite{Koposov09} note, this is a poor match to the observations.
However, in this work we adopted a more physically motivated model
where the star formation rate depends on the halo mass, resulting in
$M_* \propto \int M_{DM}^{\alpha}dt$ \citep[e.g.,][]{Juneau05, Noeske07,
Zheng07, Conroy08}, where we integrate over the time
between when the halo first crosses $\mthresh$ and the reionization
epoch.  Changing the parameter $\alpha$ directly allows us to adjust
the slope of the luminosity function and merge the two populations
into a single, continuous distribution.  In particular, the value
$\alpha = 2$, measured from high mass objects \citep{Conroy08, Drory08}
provides a close fit to the data, Figure \ref{fig:lf}.  Thus, we
conclude that a sharp cutoff of star formation can closely match the
observational data.  

\section{DISCUSSION AND CONCLUSIONS}

\subsection{Additional Constraints and Physical Processes}

Here we briefly discuss additional physics processes not addressed earlier.   One observable constraint on our model is the presence of fossil galaxies -- objects that have had no star formation since reionization.  Current observational constraints indicate that all satellites brighter than $M_V \sim -8$ have had some sort of star formation in the past 8 Gyr \citep[see the review of][and references therein]{Tolstoy09}, but that dimmer objects such as Hercules may have had all their stars formed during a single event \citep{Koch08}.  
Our base model predicts a large number of fossil galaxies.  As noted in section 2.2, there are only seven "non-fossil" galaxies, whose mass reaches our post-reionization threshold of $10^5$K.  The number of fossil galaxies in our model, however, is highly sensitive to the exact value of this threshold, while the overall abundance of satellite galaxies is not. The full impact will be explored further in a companion paper, but lowering the second cooling threshold from $10^5$K to a few $10^4$K can match the fossil constraints while not significantly impacting the predicted observed luminosity function. 

Another limitation of our model as presented is our incomplete treatment of tidal disruptions on the satellite galaxy population.  While the high resolution \VL ~simulation automatically models the impact of tidal stripping on the dark matter halo, we have ignored the further impact on the stellar population.  This is equivalent to the assumption that, once formed, a satellite galaxy remains undisrupted unless its host dark matter halos is completely disrupted, merging with the central object or dropping below the resolution limit of the simulation.  Recent studies of \cite{Maccio09} and \cite{Koposov09} have indicated that this may have a significant impact on the $z = 0$ satellite luminosity function.  To test this, we consider the criteria of \cite{Taylor04} and \cite{Maccio09} for disruption, assuming that a satellite galaxy is destroyed when the tidal mass of its host dark matter subhalo is less than $M_{bind} \equiv M(<f_{dis}r_{bind})$, where $r_{bind}$ is the radius of the subhalo, at infall within which the total energy is positive and $f_{dis}$ is a free parameter that they set to $f_{dis} = 0.1$.  \cite{Taylor04} calculate $r_{bind} = 0.77r_s$, where $r_s$ is the scale radius for an NFW profile.  In order to estimate $M_{bind}$, we adopt the mean relation for concentration given mass and redshift, $c(M,z) = 9(M_{vir}/M_*)^{-0.13}$ in a \LCDM universe from \cite{Bullock01}.  Using this, we see that, in our best fit model with $\zreion = 8$, roughly 20\% of our satellite galaxies are effected at all magnitudes.  A further investigation of this process is needed, but it appears that there may be a significant degenracy between tidal disruption and reionization redshift when attempting to match the luminosity function.  


\subsection{Future Work}

The analysis in this paper opens the door for a significant number of
future studies.  In particular, better statistics from both
simulations and observations are necessary to distinguish between the
models discussed here and confirm the result that $\zreion$ has a
significant impact on the satellite galaxy population of a Milky Way
mass halo.  

On the simulation side, we need further analysis of both large scale
and high mass-resolution simulations.  By providing excellent
statistics, large scale simulations of regions from 100--1,000 $\hinv$Mpc
will yield a significant amount of information about a wide range
of properties of Milky Way mass halos, including environmental effects on
both the reionization time and subhalo population as well as any
correlations between the two.  Clearly, there should be some
correlation between clustering or the proximity to groups and clusters
with $\zreion$ since such massive objects are the earliest sources of
photoionizing radiation.  Additionally, studies of halo assembly bias
\citep[e.g.,][]{Gao05b,Wechsler06,Hahn08} have shown correlations
between halo assembly history, clustering, and substructure
population.  It is therefore likely that these properties will also
correlate with the reionization time.  Understanding such relations
could allow us to measure the environment of the local group
to both more robustly understand the subhalo population of the Milky
Way and provide an alternate estimate for its reionization history.
Such measurements could help confirm or refute the model presented
here. 

So far we have only applied our model to simulations of two single halos -- \VL ~and the original {\em Via Lactea}.  In both simulations, we are able to reproduce the satellite luminosity and distribution functions to similar accuracy, albeit with$\zreion = 8$ for \VL ~and $\zreion = 10$ for the original simulation.  By studying additional, more highly resolved halos, we can both strengthen our predictions and accurately quantify the expected scatter in satellite galaxy population given $\zreion$ or an expected $\nsats$ for our model.  
Higher resolution will also allow us to accurately
measure the distribution of mass down to scales of 300 kpc, easing
comparisons with observations as in Figure \ref{fig:ML}.  Indeed,
these scales are just below the resolution limit in the most recent simulations
\citep[e.g.,][]{Diemand08,Springel08,Stadel08}.  
Finally, new high
resolution simulations of the detailed hydrodynamics of star formation
in the early universe can give us a better handle on how rapidly we
expect reionization to truncate star formation by more accurately
modeling the rate at which the gas in small halos is heated. Such
simulations could also give independent predictions for the values of
$\alpha$ and $\epsilon$ in equation \ref{eq:sfr}, providing another
way to test our model.

There are also a number of additional observational measurements that
would help in understanding the validity of the models presented in
this work. Most directly, deeper surveys will aid in quantifying the
abundance of substructure by both extending the luminosity function to
dimmer galaxies within the Milky Way, and providing measurements of
other systems such as M31.  Upcoming surveys such as DES, PanSTARRS, and LSST
\citep{Abbott05, Kaiser02, Ivezic08} will map the distribution of galaxies at
more than 3 dex deeper than SDSS over the entire sky, potentially
discovering hundreds of new satellites.  These improved observations
may also give us a better handle on mass measurements of the satellite
galaxies.  The depth will also allow us to potentially probe the
bright end of the satellite luminosity functions for thousands
of Milky Way mass galaxies at distances out to $\sim 60 Mpc$.
As with additional
simulations, the statistics provided by these observations will
provide significant discriminating power when applied to the models of
this work.  

Additionally, there is hope that detailed studies of the star
formation history of the local satellite galaxies can help by directly
measuring both the reionization history and rate at which reionization
quenches star formation.  Because the gradual star formation
truncation given by equation \ref{eq:fgas} results in a significant
amount of stars forming after $\zreion$, detailed modeling of star
formation may be able to discriminate between models with rapid and
gradual truncation of the star formation rate.  Unfortunately for such
modeling, any starburst activity occurring after a satellite accretes
onto the host halo would greatly confuse the results.  Because of
this, we will likely have to concentrate on the dim end of the
satellite luminosity function, restricting such studies to the Milky
Way and potentially M31.  

\subsection{Summary}

Following the recent work of \cite{Alvarez08}, we predict a broad
range of reionization times for Milky Way mass halos, ranging from
$\zreion \approx 6-12$.  We find that the time of reionization can
have a significant impact on the satellite galaxy population of a
Milky Way halo.  We investigate predictions for a simple model where,
in order to cool gas and form stars, a subhalo must reach a threshold
mass, $\mthresh$, by $\zreion$, the time it reionizes.  This model
predicts a strong dependence of the satellite galaxy population on
$\zreion$; we find that the number of satellites can vary by an order
of magnitude for a fixed $\mthresh$.  This result is in contention
with a number of previous studies which have shown minimal impact of
$\zreion$ on the satellite population.  The differing results are
due to differing assumptions about the rate at which the UV
background squelches star formation.  Previous studies have used a
gradual transition from the cold to hot gas phases, based the the work
of \cite{Gnedin00} to predict the amount of cold gas available for
star formation retained by a halo during the process of reionization.
Instead, we interpret this calculation to be more indicative of the
total amount of gas in the halo, and assume that it is rapidly heated
to a hot phase so that star formation is very quickly stopped.
Additionally, if the heating process causes a slow star formation
truncation, we believe that the quasar HeII reionization at $z \sim
4-3$ will alter the star formation history such that $\zreion$ will
more strongly impact the satellite luminosity function than previous
studies have shown.  While these two interpretations of rapid and
gradual gas heating bracket the most extreme interpretations, their
discrepancy indicates that further study into the exact heating rate
is necessary since they predict completely different dependencies on
$\zreion$.

Assuming a rapid heating of the gas from the photoionizing background,
the reionization redshift $\zreion$ impacts the ability of subhalos of
nearly all masses and radial distribution to host satellite galaxies,
with a strong impact on the satellite galaxy luminosity function.  In
spite of the simplicity of the model, the results are consistent with
both observations and detailed hydrodynamical simulations of stellar
formation.  Extrapolating the observed bright end relation between
$\vmax$ and luminosity down to satellite-sized objects, we are able to
closely reproduce the luminosity function of the Milky Way tuning only
the parameters of reionization.  This is an additional indication of
the robustness of the abundance matching method for assigning
luminosities to dark matter halos.

Because it is minimally impacted by methods for assigning magnitudes
to galaxies, we can use observations of the satellite $\vmax$ function
to place constraints on the reionization epoch of the Milky Way.  This
observed distribution is best recovered for a
reionization $\zreion = 8^{+3}_{-2}$, slightly more recent than
the ``instantaneous'' value of 11.2 from the WMAP5 data
\citep{Komatsu08}.  This value is also in excellent agreement with the
predictions from \cite{Weinmann07} for the reionization redshift of a
Milky-Way mass halo.  This model predicts that
the Milky Way should host roughly 540 satellite galaxies.  Such a
value for $\zreion$ also produces a good agreement for the radial
distribution of satellites and the luminosity function when the
abundance matching method is used.  It must still be cautioned,
however, that this result depends on assumption that the \VL
~simulation is representative of the dark matter distribution in the
Milky Way.  We get similar agreement when we adopt a model where $SFR
\propto M_{DM}^{\alpha}$, although it is important to note that this
model has a tunable parameter.  Still, it is
able to reproduce the slope of the dim end of the luminosity function
almost exactly and some late time star formation, consistent with
observations, will help relieve the (slight) tension at the bright
end. This is consistent with the recent work of
\cite{Koposov09}.  Our work lends further strength to the growing
body of research that suggests that there really is no ``missing
satellite'' problem for the Milky Way and that the next generation of
surveys may allow us to understand the entire population of local
satellite galaxies. 

\acknowledgements

MTB would like to thank B. Gerke for many helpful discussions, as well
as J. Diemand and collaborators for making the results of the 
\VL ~simulation public.  We would also like to thank Andrey Kravtsov, Oleg Gnedin, Nick Gnedin, James Bullock, Michael Kuhlen, and Charlie Conroy for many useful comments on the manuscript.  MTB and RHW also thank their collaborators
on the LasDamas project for critical input on the Gpc simulation,
which was performed using the Orange cluster at SLAC.  This work was
partially supported by NASA ATFP grant NNX08AH26G, NSF AST-0807312, and NSF AST-0908883.
RHW was supported by a Terman Fellowship at Stanford University.  LES
acknowledges support for this work by NASA through Hubble Fellowship
grant HF-01225.01 awarded by the Space Telescope Science Institute,
which is operated by the Association of Universities for Research in
Astronomy, Inc., for NASA, under contract NAS 5-26555.

\bibliography{bibliography}
\bibliographystyle{apj}

\end{document}